\documentclass[journal,twoside]{IEEEtran}

\usepackage{amsmath}
\usepackage{amssymb}
\usepackage{bm}
\usepackage{graphicx}
\usepackage{cite}

\graphicspath{{../plot/}}

\newtheorem{theorem}{Theorem}
\newtheorem{lemma}{Lemma}
\newtheorem{corollary}{Corollary}

\newcommand{\K}{\bar{K}}
\newcommand{\diff}{\mathrm{d}}

\newcommand{\rvq}{\textsf{rvq}}
\newcommand{\brvq}{\beta_{\rvq}^{\infty}}

\newcommand{\grvq}{\gamma_{\rvq}^{\infty}}
\newcommand{\irvq}{I_{\rvq}^{\infty}}
\newcommand{\bmax}{\beta_{\max}^{\infty}}
\newcommand{\bave}{\bar{\beta}}
\newcommand{\varn}{\ensuremath{\sigma_{n}^{2}}}
\newcommand{\bS}{\ensuremath{\bm{S}}}
\newcommand{\sS}{\ensuremath{\mathcal{S}}}

\newcommand{\B}{\bar{B}}
\newcommand{\bI}{\ensuremath{\bm{I}}}
\newcommand{\br}{\ensuremath{\bm{r}}}
\newcommand{\bs}{\ensuremath{\bm{s}}}
\newcommand{\bst}{\ensuremath{\tilde{\bm{s}}}}
\newcommand{\bSt}{\ensuremath{\tilde{\bm{S}}}}
\newcommand{\bn}{\ensuremath{\bm{n}}}
\newcommand{\bR}{\ensuremath{\bm{R}}}
\newcommand{\bc}{\ensuremath{\bm{c}}}
\newcommand{\bv}{\ensuremath{\bm{v}}}
\newcommand{\sV}{\ensuremath{\mathcal{V}}}
\newcommand{\bH}{\ensuremath{\bm{H}}}
\newcommand{\bA}{\ensuremath{\bm{A}}}
\newcommand{\diag}{\mathrm{diag}}

\newcommand{\me}{\mathrm{e}}

\newcommand{\sinr}{\mathrm{SINR}}
\newcommand{\irvqt}{\tilde{I}_{\rvq}^{\infty}}
\newcommand{\bAt}{\ensuremath{\tilde{\bm{A}}}}

\begin{document}

\title{Signature Quantization in Fading CDMA With Limited Feedback}

\author{Wiroonsak~Santipach,~\IEEEmembership{Member,~IEEE}%
\thanks{This work was jointly supported by Thailand Commission on
  Higher Education and the Thailand Research Fund under grant
  MRG5080385.  The material in this paper was presented in part at the
  IEEE International Conference on Communications, Beijing, China, May
  19-23, 2008.}%
\thanks{The author is with the Department of
Electrical Engineering, Faculty of Engineering, Kasetsart University,
Chatuchak, Bangkok, 10900 Thailand (email: wiroonsak.s@ku.ac.th).}}

\markboth{IEEE Transactions on Communications}{Santipach: Signature Quantization in Fading CDMA With Limited Feedback}

\maketitle

\begin{abstract}
In this work, we analyze the performance of a signature quantization
scheme for reverse-link Direct Sequence (DS)- Code Division Multiple
Access (CDMA).  Assuming perfect estimates of the channel and interference
covariance, the receiver selects the signature that minimizes
interference power or maximizes signal-to-interference plus noise
ratio (SINR) for a desired user from a signature codebook.  The
codebook index corresponding to the optimal signature is then relayed
to the user with a finite number of bits via a feedback channel.  Here
we are interested in the performance of a Random Vector Quantization
(RVQ) codebook, which contains independent isotropically distributed
vectors.  Assuming arbitrary transmit power allocation, we consider
additive white Gaussian noise (AWGN) channel first with no fading and
subsequently, with multipath fading.  We derive the corresponding SINR
in a large system limit at the output of matched filter and linear
minimum mean squared error (MMSE) receiver.  Numerical examples show
that the derived large system results give a good approximation to the
performance of finite-size system and that the MMSE receiver achieves
close to a single-user performance with only one feedback bit per
signature element.
\end{abstract}


\section{Introduction}

User performance in Direct Sequence (DS)- Code Division Multiple
Access (CDMA) depends on a signature code, which can be optimized to
increase the signal-to-interference plus noise ratio (SINR).  Several
works in the literature~\cite{rapajic95,ulukus01,wong01,rajappan02,
  viswanath99-1,viswanath99-2, rose02} have investigated a joint
transmitter-receiver signature optimization problem and have shown
that the performance difference between optimized and random signatures can be
substantial.  However, adapting the signature increases the complexity and
requires knowledge of the channel and interference covariance at both the
transmitter and receiver.  All aforementioned works assumed that
perfect estimates of the channel and interference covariance were
available.  This assumption, especially at the transmitter, is not
practical.

Typically, a receiver estimates channel coefficients and interference
covariance from pilot signals during a training period.  The accuracy
of the estimation increases with the number of available pilots.  On
the other hand, the transmitter is usually unable to directly estimate
the forward channel and may obtain channel information from the
receiver via a feedback channel.  Thus, accuracy of channel
information at the transmitter depends on the available feedback rate,
which is normally low.  In recent years, many
researchers~\cite{cdma05,dai09,love03, mukkavilli03, mimo, roh_it04,
  lau04, zhou04ciss} have proposed feedback schemes in which the
receiver computes and quantizes the optimal signature and relays the
quantized coefficients to the transmitter via a low-rate feedback
channel.  References~\cite{love03, mukkavilli03, mimo, roh_it04,lau04,
  zhou04ciss} considered multiantenna systems where spatial signatures
were optimized and quantized.  In the present paper, the focus is on
signature quantization in DS-CDMA and its performance, which depends
largely on the quantization codebook and available feedback rate.

The signature codebook is known {\em a priori} at both the transmitter
and receiver.  With $B$ feedback bits, the receiver selects the
signature vector, which maximizes the instantaneous SINR, from
$2^B$-signature codebook and relays the corresponding index to the
transmitter via an error-free feedback channel.
References~\cite{cdma05} proposed a Random Vector
Quantization (RVQ) codebook, consisting of independent isotropically
distributed vectors, and showed that the RVQ codebook was optimal
(i.e., maximize the SINR over all codebooks) in a large system limit,
in which the number of users $K$, processing gain $N$, and feedback
bits $B$, tend to infinity with fixed $\K = K/N$ and $\B = B/N$.  The
upper bound on asymptotic SINR for a single-user matched filter was
derived in \cite{cdma05}, where in addition, a minimum mean square
error (MMSE) receiver was considered and an approximation for a large
system SINR was derived.  The large system performance was shown to
predict the performance of a finite-size system well for small $\B$.

Recently, \cite{dai09} derived the exact expression of a large system
SINR for the RVQ codebook, assuming that the channel was corrupted by
additive white Gaussian noise (AWGN) and the matched filter was used
by the receiver. (Similar results for the performance of RVQ in a
multiantenna system were derived by~\cite{mimo}.)  Here we extend
those results shown by~\cite{dai09} to a multipath fading channel and
to arbitrary transmit power across users.  We apply similar techniques
to those used in ~\cite{dai09,mimo} to derive expressions for an
asymptotic SINR with a linear MMSE receiver.  For the MMSE receiver,
we first consider the AWGN channel without fading and derive the exact
expression for a large system SINR, which is a function of $\K$ and
$\B$.  We remark that the expressions for the MMSE receiver are not
trivial extensions of \cite{dai09,mimo}.  A comparison is shown
between the large system SINR and the approximation derived in
\cite{cdma05}, which overestimates the performance for large $\B$.
Numerical examples show that the performance of the finite-size system
is estimated very well by the large system results.  From the examples
shown, a linear MMSE receiver with one feedback bit per signature
element achieves close to the performance with unlimited feedback.

\section{System Model}

We consider a discrete-time reverse-link synchronous DS-CDMA in which
there are $K$ users and processing gain $N$.  The $N \times 1$
received vector is given by
\begin{equation}
  \br = \sum_{k=1}^K \sqrt{A_k} \bH_k \bs_k b_k + \bn
\end{equation}
where $\sqrt{A_k}$ is the amplitude of user $k$, $\bH_k$ is the $N
\times N$ channel matrix for user $k$, $\bs_k$ is the $N \times 1$
signature vector for user $k$, $b_k$ is the transmitted symbol for
user $k$, and $\bn$ is AWGN with zero mean and covariance $\varn \bI$.
For the AWGN channel with no fading, $\bH_k = \bI$.  For the
frequency-selective channel, we assume that the symbol duration is
much longer than the delay spread and, thus, we discard any
inter-symbol interference.  Assuming that each user traverses $L$
Rayleigh fading paths, the channel matrix is given by
\begin{equation}
  \bH_k = \left[ 
    \begin{array}{ccccccc}
      h_{k,1} & 0      & \cdots & 0      & 0      & \cdots  & 0 \\
      \vdots & h_{k,1} &        & \vdots & 0      &        & \vdots\\
      h_{k,L} & \vdots & \ddots & 0      & \vdots &        & 0\\
      0      & h_{k,L} &        & h_{k,1} & 0      & \cdots  & 0 \\
      \vdots & 0      & \ddots & \vdots & h_{k,1} &        & 0\\
      0      & \vdots &        & h_{k,L} & \vdots & \ddots & 0\\
      0      & 0      & \cdots & 0      & h_{k,L} & \cdots & h_{k,1}
    \end{array}
    \right]
\label{bHk}
\end{equation}
where fading gains for user $k$, $h_{k,1}, \ldots, h_{k,L}$, are
complex Gaussian random variables with zero means and variances
$E|h_{k,1}|^2, \ldots, E|h_{k,l}|^2$, respectively and are independent
across users and fading paths.  For a flat fading channel ($L=1$),
$\bH_k = h_{k,1} \bI$.

The receiver applies a linear filter on the received signal to obtain the
received symbol.  We consider both the matched filter and linear MMSE
receiver and assume, without loss of generality, that user 1 is the
user of interest.  The matched filter for user 1 is given by
$\bc_1 = \bst_1$
where we let $\bst_k \triangleq \bH_k\bs_k$, which is the effective
signature for user $k$.  The matched filter is simple and can be a
performance benchmark for a more complex receiver. The associated SINR
is given by
\begin{equation}
  \gamma = \frac{|\sqrt{A_1} \bc_1^{\dag} \bst_1|^2}{\bc_1^{\dag}
    \bR_1 \bc_1} = \frac{A_1 (\bs_1^{\dag} \bH_1^{\dag}\bH_1
    \bs_1)^2}{\bs_1^{\dag} \bH_1^{\dag}\bR_1 \bH_1 \bs_1}
\label{mf_sinr}
\end{equation}
where the interference-plus-noise covariance is given by
  $\bR_1 = E[\br_1 \br_1^{\dag}]$
where expectation is over transmitted symbols and noise and
\begin{equation}
  \br_1 = \br - \sqrt{A_1} \bH_1\bs_1 b_1 = \sum_{k = 2}^K \sqrt{A_k}
  \bH_k\bs_k b_k + \bn .
\label{br1}
\end{equation}
Assuming that the $b_k$'s are independent and identically distributed
 ({\em i.i.d.}) with zero mean and unit variance, we have
\begin{equation}
 \bR_1 = \sum_{k=2}^K A_k \bst_k \bst_k^{\dag} + \varn \bI = \bSt_1
 \bA_1 \bSt_1^{\dag} + \varn \bI ,
\label{R1}
\end{equation}
where $\bSt_1$ is the $N \times (K-1)$ effective signature matrix
whose columns consist of $\bst_k$, $\forall k \ne 1$ and $\bA_1$ is
the $(K-1) \times (K-1)$ diagonal matrix whose diagonal entries
are $A_2, \dots, A_K$.

Next, we consider the linear MMSE filter for user 1 given by
  $\bc_1 = \bR^{-1} \bst_1$
where the received covariance is
\begin{equation}
  \bR = E [\br \br^{\dag}] = \sum_{k=1}^K A_k \bst_k
  \bst_k^{\dag} + \varn \bI ,
\end{equation}
with the same assumption that the $b_k$'s are {\em i.i.d.} with zero mean and unit
variance.  Similar to the matched filter, we can compute the SINR for
user $1$ given by
\begin{equation}
  \beta = A_1 \bs_1^{\dag} \bH_1^{\dag} \bR^{-1}_1 \bH_1 \bs_1
 \label{sinr}
\end{equation}
where the matrix inversion lemma was used to simplify the expression.  The
linear MMSE receiver is shown to be robust in suppressing
multiple-access interference~\cite{madhow_honig}.  We note that, for
given $\bR_1$ and $\bH_1$, the SINR for user $1$ is a function of the
signature $\bs_1$ for both receivers.

The receiver, which is assumed to have a perfect estimate of the
interference covariance $\bR_1$ and channel matrix $\bH_1$, can
optimize the signature for the desired user to maximize the received
SINR. Ideally, the receiver sends the optimal signature back to user
$1$ via a feedback channel and the user changes the signature
accordingly.  Practically, a feedback channel has limited rate and
thus, the receiver can only relay a finite number of feedback bits to
the user.  (We assume that the feedback does not incur any errors.)
With $B$ bits, the receiver selects the signature from a signature set
or codebook containing $2^B$ signatures.  This codebook is designed
{\em a priori}, and is known to both the user and receiver.  The
performance of the optimized user depends on the codebook.
References~\cite{cdma05,mimo,roh_it04,narula98,mukkavilli03,love03,lau04}
proposed codebook designs and analyzed the associated performance.
(All except \cite{cdma05} were in the context of a spatial signature in a
multiantenna channel.)  In this work, we analyze the performance of a
Random Vector Quantization (RVQ) codebook proposed by
\cite{cdma05}.  We consider an RVQ codebook
\begin{equation}
  \sV = \{ \bv_1,\ldots,\bv_{2^B} \}
\end{equation}
in which the $\bv_j$'s are independent isotropically distributed with
unit norm ($\| \bv_j \| = 1$).  In other words, signature vectors in
the RVQ codebook are uniformly distributed on a surface of an
$N$-dimensional unit sphere.  In \cite{cdma05, mimo, dai09}, RVQ was
shown to maximize the SINR over all quantization codebooks in a large
system limit to be defined.  Although RVQ is optimal in the large system
limit~\cite{cdma05}, it was shown to perform close to the optimal
codebook designed for a finite-size system~\cite{commag04}.

Given the codebook $\sV$, the receiver selects
\begin{equation}
  \bs_1 = \arg \max_{\bv_j \in \sV} { \sinr ( \bv_j )} .
\end{equation}
The index of the optimal signature vector is relayed to user $1$ via a
feedback channel without delay.  A feedback delay in a time-varying
channel will degrade performance, since the optimized signature
relayed from the receiver will be outdated by the time it is used at
the transmitter.  Our model also applies to block fading in which
channel coefficients remain relatively static for a period of time to
allow meaningful feedback.  We are interested in analyzing the SINR,
which is a function of available feedback bits, for both the matched
filter and the MMSE receiver with {\em zero-delay} and {\em
  error-free} feedback.

\section{Large System Performance}

\subsection{Matched filter}

We first consider the AWGN channel without fading ($\bH_i = \bI$, for all $i$)
for which the optimal signature that maximizes SINR also minimizes the
interference.  Given the RVQ codebook $\sV$, the optimal signature is
given by
\begin{equation}
  \bs_1 = \arg \min_{\bv_j \in \sV} \, \{ I(\bv_j) \triangleq
  \bv^{\dag}_j\bR_1 \bv_j \}
\label{hat}
\end{equation}
where $I$ is the instantaneous interference power.  Since the $\bv_j$'s in
the RVQ codebook are {\em i.i.d.}, the corresponding $I(\bv_j)$'s for a
given $\bR_1$ are also {\em i.i.d.} and thus, the associated
interference averaged over the codebook is given by
\begin{multline}
  E_{\sV} [ \min \{ I(\bv_1) , \ldots, I(\bv_{2^B}) \} | \bR_1] \\
  = 2^B \int_0^\infty x [1 - G_{I | \bR_1}(x)]^{2^B -1
  }g_{I | \bR_1}(x) \, \diff x
\label{EI}
\end{multline}
where $G_{I | \bR_1}(\cdot)$ and $g_{I | \bR_1}(\cdot)$ are the
cumulative distribution function (cdf) and probability density
function (pdf) for $I(\bv_j)$, respectively.  It is difficult to
evaluate \eqref{EI} for any finite $N$, $K$, and $B$.  However,
\cite{cdma05,dai09} showed that the interference power converges to a
deterministic value in a large system limit, in which $K$, $N$, and
$B$ all tend to infinity with fixed normalized load $\K = K/N$ and
normalized feedback bits $\B = B/N$.  Applying the theory of extreme
order statistics~\cite{galambos}, similar to \cite{cdma05}, the large
system interference power with a fading channel is given by
\begin{equation}
  I^{\infty}_{\rvq} = \lim_{(N,K,B) \to \infty} G^{-1}_{I | \bR_1}
  (2^{-B})
\label{Iir}
\end{equation}
where the empirical eigenvalue distribution of $\bR_1$ converges
almost surely to a nonrandom limit as $(N,K) \to \infty$ with fixed
$K/N$.  Rearranging \eqref{Iir} gives
\begin{equation}
  \lim_{\begin{subarray}{c} (N,K,B) \to \infty \\ z \to \irvq
  \end{subarray}} [G_{I | \bR_1 }(z)]^{\frac{1}{N}} = 2^{-\B} .
\label{nkb}
\end{equation}

Reference \cite[Theorem 1]{dai09} showed that
\begin{equation}
  \lim_{\begin{subarray}{c} (N,K,B) \to \infty \\ z \to \irvq
  \end{subarray}} [G_{I | \bR_1}(z)]^{\frac{1}{N}} = 
   \exp\{ - \Psi(\rho^*,\irvq) \} \label{ccN}
\end{equation}
where
\begin{gather}
  \Psi(\rho,\irvq) = \int \log(1 + \rho(\lambda - \irvq)) g_{\bR_1}
  (\lambda) \, \diff \lambda, \label{pri} \\ \rho^* = \arg \max_{0 <
    \rho < \frac{1}{\irvq - \lambda_{\min}}}
  \Psi(\rho,\irvq), \label{arm}
\end{gather}
and $g_{\bR_1} (\cdot)$ is an asymptotic eigenvalue density for
$\bR_1$ and $\lambda_{\min}$ is the asymptotic minimum eigenvalue of
$\bR_1$.  Equating \eqref{nkb} and \eqref{ccN}, we have that $\irvq$
satisfies
\begin{equation}
   \Psi(\rho^*,\irvq) = \B \log(2) .
\label{PriB}
\end{equation}

With \eqref{R1}, Eq. \eqref{pri} becomes
\begin{align}
   &\Psi(\rho,\irvq) \nonumber\\
   &= \int \log(1 + \rho(\lambda + \varn- \irvq))
  g_{\bS_1 \bA_1 \bS_1^{\dag}} (\lambda) \, \diff \lambda .\\
    &= \log(1 + \rho \varn - \rho \irvq) + \int \log (1
  + \xi \lambda) g_{\bS_1 \bA_1 \bS_1^{\dag}} (\lambda) \, \diff \lambda\\ 
   & =
  \log(1 + \rho \varn - \rho \irvq) + \nu_{\bS_1 \bA_1 \bS_1^{\dag}} (
  \xi) \label{Sxi}
\end{align}
where
\begin{equation}
  \xi \triangleq \frac{\rho}{1 + \rho \varn - \rho \irvq} \label{xit}
\end{equation}
and $\nu_{\bS_1 \bA_1\bS_1^{\dag}} (\cdot)$ is the Shannon transform
for an asymptotic eigenvalue distribution for $\bS_1
\bA_1\bS_1^{\dag}$.  Reference \cite{tulino_verdu} defined the Shannon
transform for a density function $f_{X} (\cdot)$ as follows
\begin{equation}
   \nu_{X} (\gamma) = \int \log (1 + \gamma x)
   f_X (x) \, \diff x .
\label{nux}
\end{equation}
Suppose $\bs_k$, $2 \le k \le K$, has independent complex Gaussian
entries with zero mean and variance $1/N$ ($\| \bs_k \| \to 1$).  The
eigenvalue distribution for $\bS_1\bS^{\dag}_1$ converges to a
deterministic function as $N,K \to \infty$, with fixed $\K$
\cite{marcenko} and we assume that the empirical distribution of $A_2,
\ldots, A_K$ converges to a limit.  It was shown
by~\cite{tulino_verdu} that
\begin{multline}
  \nu_{\bS_1 \bA_1\bS_1^{\dag}} (w) = \K \nu_{\bA_1}(w \eta_{\bS_1
    \bA_1\bS_1^{\dag}} (w)) - \log(\eta_{\bS_1 \bA_1\bS_1^{\dag}} (w)) \\
  + \eta_{\bS_1 \bA_1\bS_1^{\dag}} (w) -1 
\label{w_1}
\end{multline}
where $\eta_{\bS_1 \bA_1\bS_1^{\dag}} (\cdot)$ is the $\eta$-transform
for the asymptotic eigenvalue distribution for $\bS_1
\bA_1\bS_1^{\dag}$ and the $\eta$-transform for a distribution for
random variable $X$ was defined in \cite{tulino_verdu} as follows
\begin{align}
   \eta_X(\gamma) = \int \frac{1}{1 + \gamma x} f_X (x)\, \diff x .
\label{etax}
\end{align}
With the earlier assumption on the distribution for $\bS_1$,
\cite{tulino_verdu} showed that $\eta_{\bS_1 \bA_1\bS_1^{\dag}}(x)$ is
the solution to the following fixed point equation
\begin{equation}
  \K = \frac{1 - \eta_{\bS_1\bA_1\bS_1^{\dag}}(x)}{1 - \eta_{\bA_1}(x
    \eta_{\bS_1\bA_1\bS_1^{\dag}} (x))} . \label{Kfr1}
\end{equation}

Combining \eqref{PriB}, \eqref{Sxi} \eqref{w_1}, and \eqref{Kfr1}, we
have our first main result.
\begin{theorem}
\label{mfrvq}
  The large system interference power $\irvq$ at the output of
  single-user matched filter satisfies the following equation
  \begin{multline}
    \max_{0 < \rho < \frac{1}{\irvq - \lambda_{\min}}} \{ \log(1 +
    \rho \varn - \rho \irvq) + \K \nu_{\bA_1}(\xi (\Theta(\xi))\\
    - \log\Theta(\xi) + \Theta(\xi) - 1 \} = \B \log(2) 
  \end{multline}
 where $\Theta(x)$ is the solution to the following fixed point equation
 \begin{equation}
   \K = \frac{1 - \Theta (x)}{1 - \eta_{\bA_1}(x \Theta (x))} \label{KfT}
 \end{equation}
 and $\xi$ is given by \eqref{xit}.
\end{theorem}
Solving for $\irvq$ requires numerical solution in most cases.
However, for equal power allocation ($A_1 = A_2 = \cdots = A_K$), the
explicit expression for $\irvq$ was obtained by \cite{dai09} as
follows.
\begin{corollary}[Corollary~1 in \cite{dai09}]
\label{cmf}
  Let $$\B^* = \frac{-\K \log(1 - \frac{1}{\sqrt{\K}}) -
    \sqrt{\K}}{\log(2)}$$ for $\K > 1$. For $\K > 1$ and $\B > \B^*$,
  \begin{multline}
    \irvq = \varn + (1 - \sqrt{\K})^2 \\
     + \sqrt{\K}(1 - \frac{1}{\sqrt{\K}})^{1 -
      \K} \exp(-\sqrt{\K} - \B \log(2) ). \label{ivk}
  \end{multline}
   Otherwise, $\irvq = Q + \varn$, where $Q$ satisfies the following
   equation 
  \begin{equation}
   Q = \K \me^{(Q -\K)/\K} 2^{-\B/\K}.
  \end{equation}
\end{corollary}
Thus, interference power decreases exponentially with the normalized
feedback bits and is near the single-user performance with only a few
feedback bits per processing gain.  The associated SINR for user 1 in
the large system limit is then given by
  $\grvq = \frac{1}{\irvq}$ .

If the interfering signatures are orthogonal ($\bS_1^{\dag}\bS_1 =
\bI$) with equal power allocation, the eigenvalue distribution for $\K
< 1$ is given by
\begin{equation}
  g_{\bS_1\bA_1\bS_1^{\dag}} (\lambda) = \K \delta(\lambda - 1) +
  (1-\K)\delta(\lambda)  \label{gBs}
\end{equation}
where $\delta(\cdot)$ is a Dirac delta function.  Evaluating
\eqref{PriB} with the distribution in \eqref{gBs}, we obtain the
following result.
\begin{theorem}
  The large system interference power $\irvq$ for orthogonal
  interfering signatures with equal power allocation satisfies the
  following fixed-point equation
  \begin{equation}
    (\irvq - \varn)^{\K} (1 + \varn - \irvq)^{1-\K} = \K^{\K}
    (1-\K)^{1-\K} 2^{-\B} \label{irv}
  \end{equation}
for $0 < \K < 1$.
\end{theorem}
For $\K \approx 1$, we obtain the following approximation
\begin{equation}
  \irvq \approx \varn + \K 2^{-\B} .
\end{equation}
Here, we see clearly that the interference power decreases
approximately exponentially with the normalized feedback bits when the
system has a heavy load.

For a fading channel, the signal of each user is assumed to propagate
$L$ discrete chip-spaced paths with the channel matrix for user $k$ shown
in \eqref{bHk}.  Given the RVQ codebook $\sV$, the signature that
minimizes the interference power is given by
\begin{equation}
  \bs_1 = \arg \min_{\bv_j \in \sV} \, \{ \tilde{I}(\bv_j) \triangleq
  \bv^{\dag}_j\bH^{\dag}_1 \bR_1 \bH \bv_j \} .
\end{equation}
We remark that $\bs_1$ may not maximize the SINR in \eqref{mf_sinr}.
Similar to the channel with no fading, the large system interference
power with the RVQ codebook with a fading channel is given by
\begin{equation}
  \irvqt = \lim_{(N,K,B,L) \to \infty} \min_{\bv_j \in \sV}
  \tilde{I}(\bv_j), 
\end{equation}
assuming that a sum of fading variances of each user is finite,
\begin{equation}
  \alpha_k \triangleq \sum_{l = 1}^L E|h_{k,l}|^2 < \infty \qquad
  \text{for} \quad 1 \le k \le K .
\end{equation}
Applying Theorem~\ref{mfrvq}, $\irvqt$ is determined by solving
\eqref{pri} with the asymptotic eigenvalue density $g_{\bR_1} (\cdot)$
replaced with $g_{\bH^{\dag}_1 \bR_1 \bH_1}(\cdot)$.

To determine the asymptotic eigenvalue distribution of $\bH^{\dag}_1
\bR_1 \bH_1$, we first consider $\bR_1 \bH_1 \bH^{\dag}_1$ whose
nonzero eigenvalues are the same as those of $\bH^{\dag}_1 \bR_1
\bH_1$.  Since $\bH_1$ is an $N \times N$ Toeplitz matrix, the $n$th
eigenvalue of $\bH_1$ is given by \cite{Gray72}
\begin{equation}
  \lambda_n (\bH_1) = \sum_{l=1}^L h_{1,l} \me^{j2\pi l \frac{n-1}{N}}
\end{equation}
and thus, as $(N, L) \to \infty$, 
\begin{equation}
  \lambda_n (\bH_1^{\dag}\bH_1) = \left| \sum_{l=1}^L h_{1,l}
  \me^{j2\pi l \frac{n-1}{N}} \right|^2 \longrightarrow \alpha_1
\label{ldn}
\end{equation}
almost surely.  The limit in \eqref{ldn} follows from the law of large
numbers and earlier assumptions that $h_{k,l}$ is independent across a
fading path $l$, and $\alpha_1 < \infty$.  Since all eigenvalues of
$\bH_1^{\dag}\bH_1$ converge to the same limit, the asymptotic
eigenvalue distribution of $\bH^{\dag}_1 \bR_1 \bH_1$ equals that of
$\alpha_1 \bR_1$ where $\bR_1$ is given in \eqref{R1}.

To obtain the $\eta$-transform of $g_{\bSt_1 \bA_1 \bSt_1^{\dag}}
(\cdot)$, we solve the following fixed-point
equation~\cite{tulino_verdu}
\begin{equation}
  \K = \frac{1 - \eta_{\bSt_1\bA_1\bSt_1^{\dag}}(x)}{1 - \eta_{\bAt_1}(x
    \eta_{\bSt_1\bA_1\bSt_1^{\dag}} (x))}  \label{Kfr2}
\end{equation}
where $\eta_{\bAt_1}$ is the $\eta$-transform of the asymptotic
eigenvalue distribution of
\begin{equation}
  \bAt_1 = \diag\{\alpha_2 A_2, \ldots, \alpha_K A_K\}.
\end{equation}
Comparing \eqref{Kfr2} with \eqref{Kfr1}, we deduce that the
asymptotic eigenvalue distributions of $\bSt_1\bA_1\bSt_1^{\dag}$ and
$\bS_1\bAt_1\bS_1^{\dag}$ are the same. In other words, a multipath
interferer is asymptotically equivalent to a single-path interferer
with a combined fading gain of $\alpha_k$.

Thus, the asymptotic minimum interference power with fading is given
by
\begin{equation}
  \irvqt = \alpha_1 \irvq 
\label{irvqt}
\end{equation}
where $\irvq$ is obtained by Theorem~\ref{mfrvq} with asymptotic
eigenvalue density $g_{\bS_1\bAt_1\bS_1^{\dag}} (\cdot)$ instead.
As $N \to \infty$,
\begin{equation}
  \bs_1^{\dag} \bH_1^{\dag} \bH_1 \bs_1 \to \alpha_1 .
\label{bsH}
\end{equation}
 Substituting \eqref{irvqt} and \eqref{bsH} in \eqref{mf_sinr} gives
 the associated SINR at the output of the matched filter with a fading
 channel
\begin{equation}
  \grvq = \frac{A_1 \alpha_1^2}{\irvqt} = \frac{A_1 \alpha_1}{\irvq}.
\end{equation}

\subsection{Linear MMSE Receiver}

The SINR with the optimal signature averaged over the RVQ codebook is
given by
\begin{multline}
  E_{\sV}[\max \{\beta(\bv_1),\ldots, \beta(\bv_{2^B})\} | \bR_1, \bH_1
   ]\\
   = 2^B \int_0^\infty x [ F_{\beta | \bR_1, \bH_1}(x)]^{2^B-1}
   f_{\beta | \bR_1, \bH_1}(x) \, \diff x 
\label{esV}
\end{multline}
where $f_{\beta | \bR_1, \bH_1}( \cdot )$ and $F_{\beta | \bR_1,
  \bH_1}( \cdot )$ be pdf and cdf for the output SINR $\beta(\bv_j)$,
respectively.  Similar to the matched filter, computing \eqref{esV}
for finite parameters is difficult.  Taking the large system limit as
$N, K, B \to \infty$ with fixed ratios, the SINR converges to a
deterministic value
\begin{align}
  \beta_{\rvq}^{\infty} &= \lim_{(N,K,B) \to \infty} E_{\sV} [ \max
    \{ \beta_1, \ldots, \beta_{2^B} \} | \bR_1, \bH_1]\\ &=
  \lim_{(N,K,B) \to \infty} F^{-1}_{\beta | \bR_1, \bH_1} (1 -
  2^{-B}),
\label{brvq}
\end{align}
which can be shown by applying the theory of extreme order
statistics~\cite{galambos}.  Reference~\cite{cdma05} derived the
approximation for $\beta_{\rvq}^{\infty}$ by approximating cdf for
$\beta (\bv_j$) to be Gaussian.  The approximation is a function of
$\K$, $\B$, and $\varn$ and is good for small $\B$.  For large $\B$,
it overestimates the actual performance.  In this section, we derive
exact expressions for $\beta_{\rvq}^{\infty}$.

We first consider the ideal channel with no fading ($\bH_k = \bI,
\forall k$).  We rearrange \eqref{brvq} to obtain
\begin{equation}
  \lim_{\begin{subarray}{c} (N,K,B) \to \infty \\ z \to \brvq
  \end{subarray}} [1 - F_{\beta | \bR_1}(z)]^{\frac{1}{N}} = 2^{-\B} .
\label{lbs}
\end{equation}
Similar to \cite[eq. (152)]{mimo} and \cite[Theorem 1]{dai09}, it can
be shown that
\begin{equation}
  \lim_{\begin{subarray}{c} (N,K,B) \to \infty \\ z \to \brvq
  \end{subarray}} [1 - F_{\beta | \bR_1}(z)]^{\frac{1}{N}} = 
  \exp\{ -\Phi(\rho^*, \brvq)\}
\label{sar}
\end{equation}
where
\begin{gather}
  \Phi(\rho, \brvq) = \int \log(1 + \rho ( \brvq - \frac{A_1}{\tau +
    \varn})) f_{\bS_1 \bA_1 \bS_1^{\dag}} (\tau) \ \diff \tau ,
\label{phb}\\
  \rho^* = \arg \max_{0 < \rho < \frac{1}{\bmax - \brvq}} \Phi(\rho,
  \brvq) , 
\end{gather}
$f_{\bS_1 \bA_1 \bS_1^{\dag}}(\cdot)$ is the asymptotic eigenvalue
density for $\bS_1 \bA_1 \bS_1^{\dag}$, $\bS_1$ is the $N \times (K -
1)$ signature matrix whose columns are $\bs_2, \ldots, \bs_K$, and
$\bmax$ is the asymptotic maximum eigenvalue of $A_1 \bR^{-1}_1$ and
corresponds to the SINR with infinite feedback ($\B \to \infty$).

Combining \eqref{lbs} and \eqref{sar}, $\brvq$ satisfies the following
fixed-point equation
\begin{equation}
  \Phi(\rho^*, \brvq) = \B \log(2) .
\label{phq}
\end{equation}
To evaluate $\Phi(\rho^*, \brvq)$, we rewrite \eqref{phb} as follows
\begin{align}
  &\Phi(\rho, \brvq) \\
&= \log(1 + \rho (\brvq - \frac{A_1}{\varn})) +\int \log( 1 + \zeta \tau) 
   f_{\bS_1 \bA_1\bS_1^{\dag}}(\tau) \, \diff \tau \nonumber\\
 &\quad - \int \log( 1 + \frac{1}{\varn} \tau)
  f_{\bS_1 \bA_1 \bS_1^{\dag}}(\tau) \, \diff \tau \label{ilS}\\ 
  & = \nu_{\bS_1 \bA_1\bS_1^{\dag}} (\zeta)
  -\nu_{\bS_1 \bA_1\bS_1^{\dag}} (1/\varn)+ \log(1 + \rho (\brvq -
  \frac{A_1}{\varn})) \label{nubS}
\end{align}
where 
\begin{equation}
  \zeta \triangleq \frac{1 + \rho \brvq}{\varn + \rho \brvq \varn -
    \rho A_1}
\label{zt1}
\end{equation}
and $\nu_{\bS_1 \bA_1\bS_1^{\dag}} (\cdot)$ is the Shannon transform
for the asymptotic eigenvalue distribution for $\bS_1 \bA_1\bS_1^{\dag}$.
With similar steps used to derive Theorem~\ref{mfrvq}, we obtain the
following theorem.
\begin{theorem}
\label{mmservq}
  For $\B$, the large system SINR $\brvq$ is given by
 \begin{multline}
   \max_{0 < \rho < \frac{1}{\bmax - \brvq}} \{ \log(1 + \rho (\brvq -
   \frac{A_1}{\varn})) + \K \nu_{\bA_1} (\zeta \Theta(\zeta)) \\
  - \K\nu_{\bA_1} (\sigma^{-2}_n \Theta(\sigma^{-2}_n))
   -\log(\Theta(\zeta)) + \log(\Theta (\sigma^{-2}_n))\\
    + \Theta(\zeta)
   - \Theta (\sigma^{-2}_n) \} =\B \log(2)
\label{ptt}
\end{multline}
 where $\zeta$ and $\Theta (x)$ are given by \eqref{zt1} and
 \eqref{KfT}, respectively.
\end{theorem}

For an equal-power ($A_1 = A_2 = \ldots = A_K$) system, we can
simplify the expression for $\brvq$ as follows.
\begin{corollary}
\label{asym_sinr}
  We assume no fading, {\em i.i.d.} interfering signatures, and
  equal transmitted power across users.  For $\K \le 1$, $\brvq$
  satisfies the following equation
  \begin{multline}
    \log(\frac{\K}{1 - \brvq \varn} - \frac{1}{\brvq}) + (1 - \K)
    \log( \frac{p}{\varn}) \\ + \K \log(\frac{w(p)}{w(\varn)})
    - (1-\K)\log(\frac{1-v(p)}{1-v(\varn)}) 
     \\ - v(p) + v(\varn)
    = \B \log(2)
    \end{multline}
where
\begin{align}
  w(x) &= \frac{1}{2} (1 + \K +x + \sqrt{(1+\K+x)^2 - 4\K})\\
  v(x) &= \frac{1}{2}(1 + \K +x - \sqrt{(1+\K+x)^2 - 4\K})
\end{align}
and
\begin{equation}
  p = \frac{1 - \brvq \varn}{\K \brvq -1 + \brvq \varn} -
  \frac{1}{\brvq} + \varn .
\end{equation}

For $\K > 1$ and $\B \le \B^*$, $\brvq$ satisfies the following
equation
\begin{multline}
  \log(\frac{\K}{1 - \brvq \varn} -
  \frac{1}{\brvq}) + \log(\frac{w(p)}{w(\varn)}) \\
  - (\K-1)\log(\frac{\K
  - v(p)}{\K-v(\varn)}) - v(p) + v(\varn)
  = \B \log(2)
\end{multline}
where 
\begin{multline}
  \B^* = \frac{1}{\log(2)}( \log(\K-\sqrt{\K}+\varn) +
  \K\log(\sqrt{\K}) \\ 
  -\K\log(\sqrt{\K}-1) - \sqrt{\K} 
  - \log(w(\varn)) \\
  + (\K-1) \log(1 - \frac{v(\varn)}{\K}) 
  + v(\varn) ) .
\end{multline}

For $\K > 1$ and $\B > \B^*$,
\begin{multline}
    \brvq = \bmax(1 - 2^{-\B} [\exp\{\frac{1}{2}\K\log(\K) \\
   - (\K-1)
  \log(\frac{\K\sqrt{\K} - \K}{\K - v(\varn)}) - \log(w(\varn)) \\
+
  v(\varn) - \sqrt{\K} \}]) .
\end{multline}
\end{corollary}
The proof is shown in the Appendix. 

We can also derive the SINR when the interfering signatures are
orthogonal.  Substituting the corresponding eigenvalue distribution
\eqref{gBs} into \eqref{ilS} and simplifying give the following result.
\begin{theorem}
\label{mmservq_uni}
  For an orthogonal set of interfering signatures with $0< \K < 1$ and
  equal power allocation, the large system SINR $\brvq$ is the
  solution of the following fixed-point equation
  \begin{multline}
    (A_1 - \brvq\varn)^{\K} (\brvq - (A_1 - \brvq\varn))^{1-\K} \\
   =
    \left( \frac{A_1\K}{1 + \varn}\right)^{\K}
    \left(\frac{A_1(1-\K)}{\varn}\right)^{1-\K} 2^{-\B}.
  \end{multline}
\end{theorem}
For a system with heavy load ($\K \approx 1$), we have
\begin{equation}
  \brvq \approx \frac{A_1}{\varn} - \frac{A_1\K}{\varn(1+\varn)}
  2^{-\B}. \label{bax}
\end{equation}
The first term on the right-hand side of \eqref{bax} is the
single-user performance and thus, the performance with the MMSE
receiver also increases exponentially with $\B$, which is the same as
performance with the matched filter.

Similar to the matched-filter case, a multipath fading is
asymptotically equivalent to a single-path fading with combined gain
of $\alpha_k < \infty$.  The SINR with fading channel can be obtained
from Theorem~\ref{mmservq}, in which we replace $A_1$ with $\alpha_1
A_1$, and $\bA_1$ with $\bAt_1$.

\section{Numerical Results}
\label{num}

\begin{figure}
  \centering
  \includegraphics[width=3.3in]{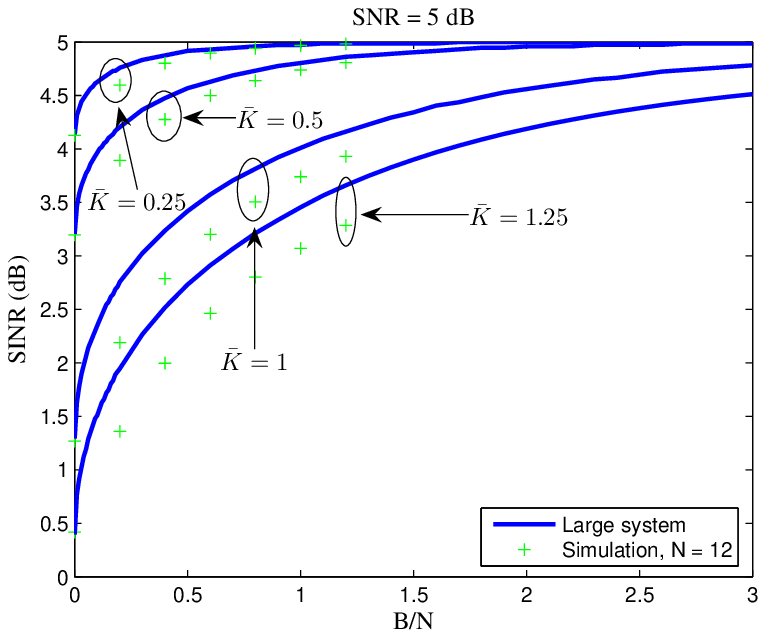} 
  \caption{Large system SINR for MMSE receiver versus
    normalized feedback bit $\B$ with different normalized loads $\K =
    0.25, 0.5, 1, 1.25$ and $\text{SNR} = 5 \ \text{dB}$.}
  \label{load}
\end{figure}

Fig.~\ref{load} shows the asymptotic SINR for the MMSE receiver in
Corollary~\ref{asym_sinr} versus a normalized feedback bit $\B$ with
different normalized loads $\K = 0.25,0.5,1,1.25$.  As expected, the
SINR increases with normalized feedback and decreases with normalized
load.  For $\K = 0.25$, RVQ achieves close to the single-user
performance with approximately $\B=0.5$ ($0.5$ bits per processing
gain or degree of freedom).  As the number of interfering users
increases, the amount of feedback required also increases to achieve a
target SINR.  For example, $\B = 3$ is needed for a system with $\K = 1$
to achieve close to the single-user performance.  We also compare the
asymptotic results with simulation results marked by the plus signs in
Fig.~\ref{load}.  We note that the large system results predict the
performance of finite-size systems ($N=12$) well.  As $N$ increases,
the gap between the simulation and analytical results is expected to
close.  The RVQ codebook requires an exhaustive search to locate the
optimal signature.  The search complexity increases exponentially with
feedback bits $B$. (For $\B = 3$, the number of entries in the RVQ codebook is
$2^{36}$.)  Thus, we do not have simulation results for a large $B$.

\begin{figure}
  \centering
  \includegraphics[width=3.3in]{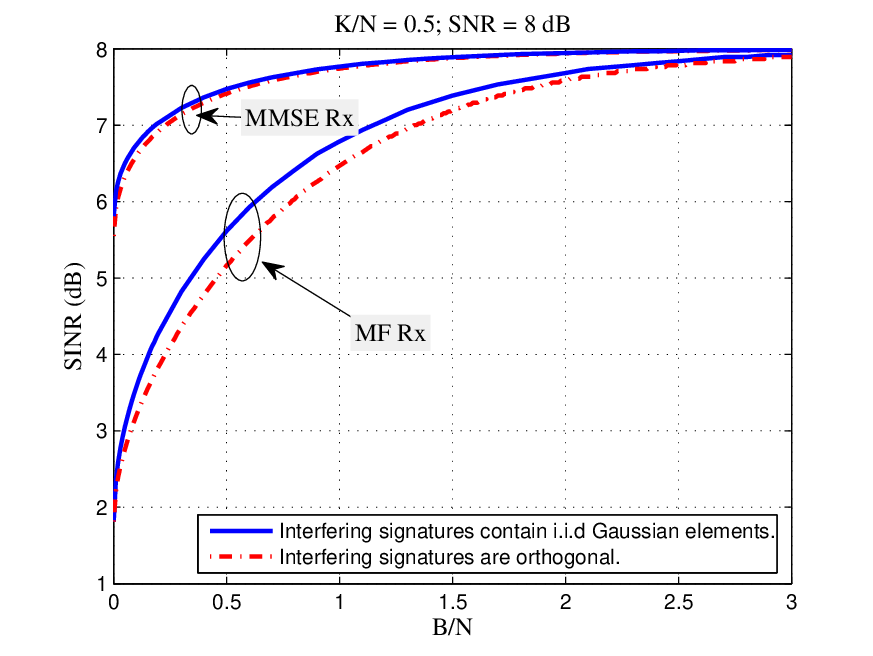} 
  \caption{Large system SINR's for different distributions
    of interfering signatures with $\K = 0.5$ and $\text{SNR} = 8
    \ \text{dB}$.}
  \label{comp_dist}
\end{figure}

Fig.~\ref{comp_dist} shows the large system performance of both
receivers, which is obtained from
Theorems~\ref{mfrvq}-\ref{mmservq_uni}, with different distributions
for interfering signatures.  We consider both sets of orthogonal and
{\em i.i.d.} Gaussian interfering signatures.  From Fig.~\ref{comp_dist}, the
system with independent Gaussian signatures performs a little better than
that with orthogonal signatures for both linear receivers.  The
difference is more pronounced with the matched filter.  Which
distribution for interfering signatures gives the maximum performance
is an interesting open problem.

\begin{figure}
  \centering
  \includegraphics[width=3.3in]{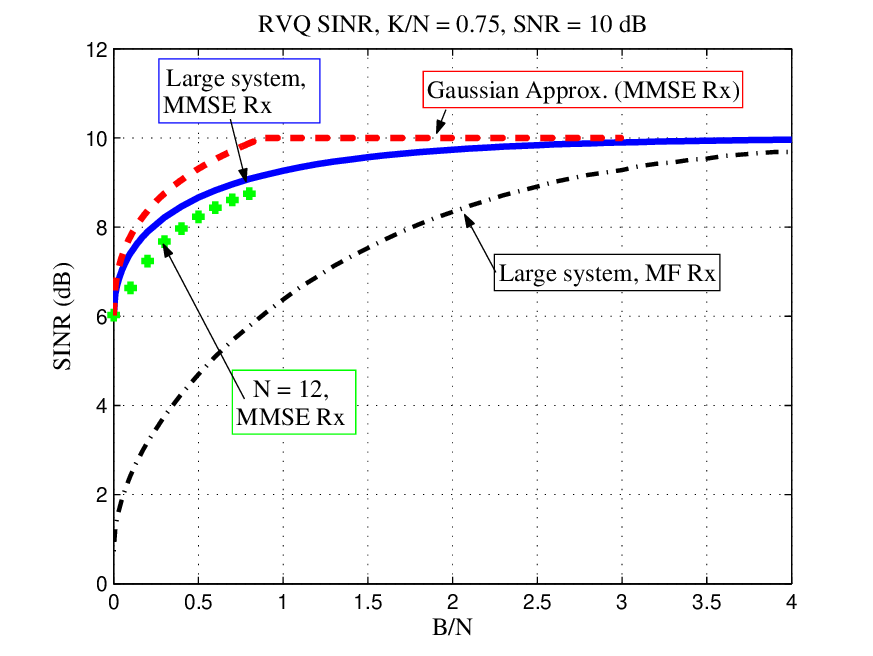} 
  \caption{Large system SINR for MMSE receiver compared with
    the approximation derived in \cite{cdma05} and the large system
    SINR for a matched filter~\cite{dai09}.  Also shown is the
    simulation result for $N = 12$, $\K = 0.75$ and $\text{SNR} = 10 \
    \text{dB}$. }
  \label{comp_sinr}
\end{figure}

In Fig.~\ref{comp_sinr}, we compare the asymptotic SINR for the MMSE
receiver in Corollary~\ref{asym_sinr} with the approximation derived
in \cite{cdma05} for $\K = 0.75$ and $\text{SNR} = 10 \ \text{dB}$.
Also shown are the simulation results with $N = 12$.  The large system
SINR is closer to the simulated performance than the approximation.
We also show the RVQ performance of the matched filter in
Corollary~\ref{cmf} ~\cite{dai09} with that of the MMSE receiver.  The
performance difference can be substantial for small to moderate $\B$.
With $1$ feedback bit per degree of freedom, the MMSE receiver
outperforms a matched filter by as much as $30 \%$.  However, the MMSE
filter is more complex than the matched filter.  Therefore, there is a
performance tradeoff between the feedback and receiver complexity.

\begin{figure}
  \centering
  \includegraphics[width=3.3in]{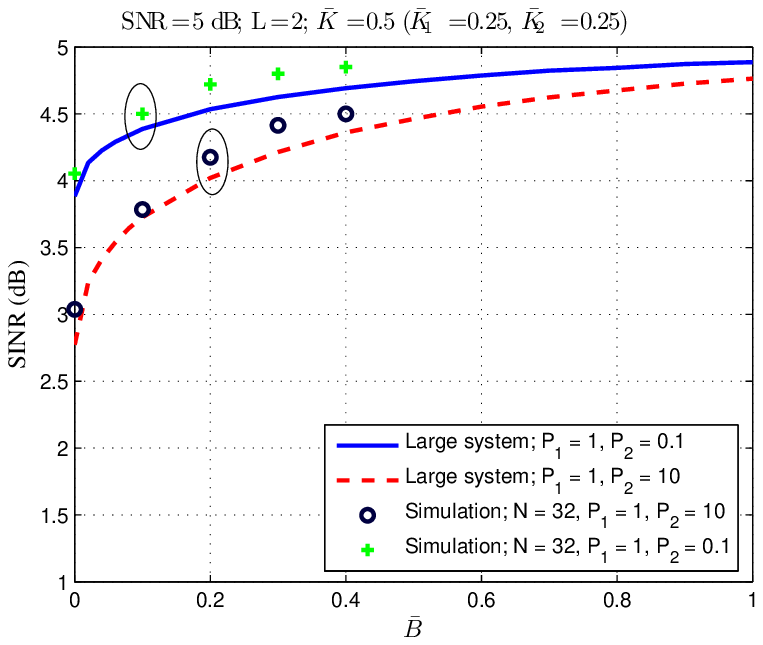} 
  \caption{Large system SINR for MMSE receiver and multipath fading
    with two groups of users with simulation results. $\text{SNR} = 5
    \, \text{dB}$, number of paths $L = 2$ for all users, and $\K =
    0.5$.}
  \label{fade}
\end{figure}

We also simulated a multipath fading channel, in which each user's
signal traverses two paths with different gains ($E | h_{k,1}|^2 =
0.9$ and $E | h_{k,2}|^2 =0.1, \forall k$).  Furthermore, $K$
interfering users are divided into two groups. $K_1$ users transmit
signal with $A_k = P_1$ while $K_2$ users transmit with $A_k = P_2$.  This
scenario may follow from a system with differentiated quality of
service.  We obtain the large system SINR from Theorem~\ref{mmservq}
with the asymptotic distribution of $\bA_1$
\begin{equation}
  f_{\bA_1} ( a ) = \frac{\K_1}{\K}\delta(a-P_1) + \frac{\K_2}{\K}
  \delta(a-P_2)
\end{equation}
where normalized loads $\K_1 = K_1/N$ and $\K_2 = K_2/N$.  Both the
large system and corresponding simulated results with $\K_1 = \K_2 =
0.25$ and different sets of $P_1$ and $P_2$ are shown in
Fig.~\ref{fade}.  The large system performance approximates closely
the performance of the system with $N=32$. As $N$ grows, the
performance of a finite-size system will converge to that of the large
system.  In this example, reducing the transmit power of one group of
users by 20 dB ($P_2$ from 10 to 0.1) decreases the required feedback
to achieve 0.5 dB away from the single-user performance by $\B = 0.4$.

Consider a reverse-link channel, in which signals from users attenuate
with distance.  The received signal power at distance $d$ away from
the transmitter is given by~\cite{rappaport}
  \begin{equation}
    P_r = P_t K_p \left( \frac{d_0}{d} \right)^\tau
  \end{equation}
where $P_t$ is the transmitted power, $K_p$ is the constant that depends
on antenna characteristics and average channel attenuation, $d_0$ is a
reference distance, and $\tau$ is the path-loss exponent.  We assume
that interfering users are placed uniformly in a circular cell with
distance $d_k$ away from the base station, where $d_0 \le d_k \le
d_{\max}, \forall k \ne 1$ and that interfering users are transmitting
with the same power $P_t$.  With the path-loss exponent $\tau = 2$, a
probability density for received power of interfering users at the
base station is given by
\begin{equation}
  f_{\bA_1} (a) = \frac{d_0^2}{K_p P_t (d_{\max}^2 - d_0^2)}
  \left( \frac{K_p P_t}{a} \right)^{2}
\label{fba}
\end{equation}
where
\begin{equation}
  K_p P_t \left( \frac{d_0}{d_{\max}} \right)^2 \le a \le K_p P_t .
\end{equation}

\begin{figure}
  \centering
  \includegraphics[width=3.3in]{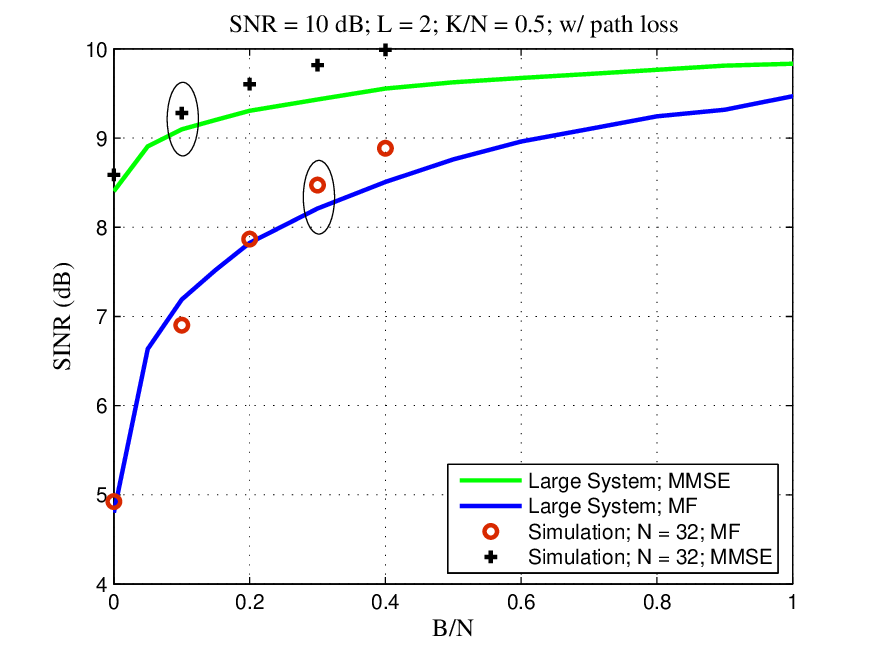} 
  \caption{A large system SINR for matched filter and MMSE receiver
    with multipath fading and path loss with simulation
    results. $\text{SNR} = 10 \, \text{dB}$, number of paths $L = 2$
    for all users, $\K = 0.5$, path-loss exponent $\tau = 2$.}
  \label{pathloss_fade}
\end{figure}

We also assume that all users experience 2-path fading with combined
gain $\alpha_k = 1, \forall k$. Thus, $f_{\bAt_1} = f_{\bA_1}$.
Substituting \eqref{fba} in \eqref{etax} and \eqref{nux} gives
 \begin{equation}
   \eta_{\bAt_1}(x) = 1 - \frac{d_0^2 K_p P_t}{d_{\max}^2 - d_0^2} x
   \log \left(\frac{x + \frac{d_{\max}^2}{K_p P_t d_0^2}}{x +
     \frac{1}{K_p P_t}} \right)
 \end{equation}
and 
\begin{multline}
  \nu_{\bAt_1}(x) = \frac{d_0^2 K_p P_t}{d_{\max}^2 - d_0^2} \bigg[
    \frac{d_{\max}^2}{K_p P_t d_0^2} \log \left( 1 + x K_p P_t
    \frac{d_0^2}{d_{\max}^2}\right) \\
    - \frac{1}{K_p P_t}\log (1 + x K_p
    P_t) \bigg] - \eta_{\bAt_1}(x) + 1 .
\end{multline}
Applying Theorems~\ref{mfrvq} and \ref{mmservq}, we obtain the large
system performance for matched-filter and MMSE receivers.
Fig.~\ref{pathloss_fade} compares the large system results shown in
solid lines with simulation results shown with markers for $P_t = 1$,
$K_p = 1$, $A_1 = 0.1$, $d_0 = 0.1$, $d_{\max} = 1$, and $L = 2$.  The
large system results approximate the simulation results for a system
with $N = 32$ well and we expect the difference between the two to
narrow as the system size increases.  With one feedback bit per signature
element, both receivers perform close to the single-user performance.

\section{Conclusions}

We have derived expressions for the large system SINR for RVQ with
both a matched filter and linear MMSE receiver.  The SINR is a
function of a normalized load (number of users per degree of freedom)
and a normalized feedback bit (number of feedback bits per degree of
freedom).  Both the AWGN channel with no fading and the multipath
fading channel with arbitrary transmit power allocation were
considered.  The SINR of the quantized signature for both receivers
increased approximately exponentially with $\B$.  For a small load,
RVQ achieved close to the single-user performance with only a fraction
of the feedback bit per quantized signature coefficient.  The
performance of the MMSE receiver was compared with that of a matched
filter derived in \cite{dai09} and it was shown that the performance
gap was large for a small $\B$.  The simpler matched filter requires
more feedback to achieve a target SINR than the MMSE receiver does.

This work assumed that the receiver could estimate the channel and
interference covariances perfectly.  In practice, a very accurate
channel estimation is achieved by a large amount of training.  How the
performance of RVQ is affected by an imperfect channel estimate at the
receiver (or limited training) was studied by~\cite{train10}.
This present work considered signature quantization for a {\em single}
user.  Future work may include performance analysis of a group of
users with RVQ-quantized signatures.

\appendix[Proof of Corollary~\ref{asym_sinr}]

We rewrite \eqref{phb} as follows
\begin{align}
  &\Phi(\rho, \brvq) \nonumber\\
  &= \int \log(1 + \rho ( \brvq - \frac{1}{x +
    \varn} )) f_{\bS_1\bS_1^{\dag}} (x) \ \diff x \\ & = \int \log(x +
  \varn + \rho(\brvq(x + \varn) - 1)) f_{\bS_1 \bS_1^{\dag}}(x)
  \ \diff x \nonumber\\
  &\quad - \int \log(x + \varn) f_{\bS_1 \bS_1^{\dag}}(x) \ \diff x
\label{pdx}
\end{align}
where
\begin{equation}
  f_{\bS_1 \bS_1^{\dag}}(x) = \frac{\sqrt{(x - a)(b - x)}}{2 \pi x} 
   \qquad \text{for} \quad
  a \le x \le b ,
\end{equation}
where $a = (1 - \sqrt{\K})^2$ and $b = (1 + \sqrt{\K})^2$ for $\K > 1$.

To determine $\rho^*$, we take the first derivative of \eqref{pdx}
with respect to $\rho$ given by
\begin{multline}
  \frac{\diff \Phi(\rho, \brvq)}{\diff \rho} = \frac{1}{\rho} -
  \frac{1}{\rho ( \rho \brvq + 1)} \\
  - \frac{1}{(\rho \brvq + 1)^2}
  \underbrace{\int \frac{1}{x - y} f_{\bS_1 \bS_1^{\dag}}(x) \ \diff x}_{\sS_f(y)}
\label{fdp}
\end{multline}
where $\sS_f(\cdot)$ is the Stieltjes transform of $f_{\bS_1
  \bS_1^{\dag}}(\cdot)$ and
\begin{equation}
  y \triangleq \frac{\rho}{\rho \brvq + 1} - \varn .
\label{ytf}
\end{equation}
We solve for $\rho^*$ (or equivalently $y^*$) by setting \eqref{fdp}
to zero and obtain
\begin{equation}
  \sS_f(y^*)  = (\rho^* \brvq + 1) \brvq .
\label{sSg}
\end{equation}
Substituting the Stieltjes transform of $f_{\bS_1
\bS_1^{\dag}}(\cdot)$ and using the change of variable from
\eqref{ytf} in \eqref{sSg} give
\begin{multline}
  \frac{-1 + \K - y^* \pm \sqrt{(y^*)^2 - 2(\K + 1)y^* + (\K-1)^2}}{2
  y^*} \\
 = \frac{\brvq}{1 - \brvq(y^* + \varn)} .
\label{f1k}
\end{multline}
Simplifying \eqref{f1k} gives
\begin{equation}
  y^* = \frac{(1 - \brvq(\K-1+\varn))(1 - \brvq \varn)}{\brvq(1 -
  \brvq(\K + \varn))} .
\end{equation}
With a change of variable \eqref{ytf}, we obtain
\begin{equation}
  \rho^* = \frac{\K}{\brvq(1 - \brvq \varn)} - \frac{1}{(\brvq)^2} -
  \frac{1}{\brvq} .
\label{rfa}
\end{equation}
To show that $\rho^*$ achieves the maximum, we prove that $\Phi(\rho,
\brvq)$ is concave down when $\rho = \rho^*$ by computing the second
derivative of $\Phi(\rho,\brvq)$ in \eqref{pdx} with respect to $\rho$
\begin{align}
  &\frac{\diff^2 \Phi(\rho, \brvq)}{\diff \rho^2} \nonumber\\
  &= - \int_a^b
  \frac{(\brvq(x + \varn)-1)^2}{(x + \varn + \rho(\brvq(x +
  \varn)-1))^2} f_{\bS_1 \bS_1^{\dag}}(x) \ \diff x \\
   & \le 0 .
\end{align}

For large enough $\brvq \ge {\brvq}^*$, $\rho^*$ in \eqref{rfa} can
exceed $1/(\bmax - \brvq)$.  To determine ${\brvq}^*$, we set
\begin{equation}
  \frac{\K}{\brvq(1 - \brvq \varn)} - \frac{1}{(\brvq)^2} -
  \frac{1}{\brvq} = \frac{1}{\bmax - \brvq} .
\label{fKb}
\end{equation}
Simplifying \eqref{fKb} gives the following quadratic equation
\begin{equation}
  [\K + \varn -\bave \varn] {\brvq}^{2} + [(1 - \K - \varn)\bave -
  1] {\brvq} + \bmax = 0 .
\label{Kvn}
\end{equation}
Solving \eqref{Kvn} gives the only solution 
\begin{equation}
  {\brvq}^* = \frac{\K - \sqrt{\K} + \varn}{(\K-\sqrt{\K})^2 +
  2\varn(\K - \sqrt{\K}) + \sigma^4_n} .
\label{bst}
\end{equation}
Thus,
\begin{equation}
  \rho^* = \left\{ \begin{array}{l@{,\quad}l} \frac{\K}{\brvq(1 -
            \brvq \varn)} - \frac{1}{(\brvq)^2} - \frac{1}{\brvq} &
            \bave \le \brvq \le {\brvq}^* \\ 
            \frac{1}{\bmax - \brvq} & \brvq > {\brvq}^*
           \end{array}
           \right. .
\label{rhs}
\end{equation}

Substituting $\rho = \rho^*$ in \eqref{pdx} and rearranging the
equation give
\begin{multline}
  \Phi(\rho^*,\brvq) = \log(\rho^* \brvq + 1) \\
  + \int^b_a \log(x + \varn - \frac{\rho^*}{\rho^* \brvq + 1}) f_{\bS_1 \bS_1^{\dag}}(x) \ \diff x \\
  - \int_a^b \log(x + \varn) f_{\bS_1 \bS_1^{\dag}} (x) \ \diff x .
\label{phr}
\end{multline}
First, we consider the case where $\bave \le \brvq \le {\brvq}^*$.
Substituting $\rho^*$ into the first term in \eqref{phr} gives
\begin{equation}
  \log(\rho^* \brvq + 1) = \log(\frac{\K}{1 - \brvq \varn} -
  \frac{1}{\brvq}) .
\label{lrb}
\end{equation}

To evaluate the two integrals in \eqref{phr}, we apply the following
lemma.
\begin{lemma}[Eqs. (6)--(8) in \cite{rapajic00}]
\label{lm}
  For $\K \ge 1$,
  \begin{multline}
    \int^b_a \log(x + \alpha) f_{\bS_1 \bS_1^{\dag}}(x) \ \diff x \\
    = \log(w(\alpha)) -
    (\K-1)\log(1 -\frac{1}{\K} v(\alpha)) - v(\alpha)
  \end{multline}
where
\begin{align}
  w(\alpha) &= \frac{1}{2} (1 + \K +\alpha + \sqrt{(1+\K+\alpha)^2 - 4\K}) ,\\
  v(\alpha) &= \frac{1}{2}(1 + \K +\alpha - \sqrt{(1+\K+\alpha)^2 - 4\K}) .
\end{align}
\end{lemma}

Using Lemma~\ref{lm} and \eqref{lrb}, we can evaluate \eqref{phr} for
$\bave \le \brvq \le {\brvq}^*$,
\begin{multline}
  \Phi(\rho^*,\brvq) = \log(\frac{\K}{1 - \brvq \varn} -
  \frac{1}{\brvq}) + \log(\frac{w(p)}{w(\varn)}) \\
  - (\K-1)\log(\frac{\K
  - v(p)}{\K-v(\varn)}) - v(p) + v(\varn)
\end{multline}
where 
\begin{equation}
  p = \frac{1 - \brvq \varn}{\K \brvq -1 + \brvq \varn} -
  \frac{1}{\brvq} + \varn .
\end{equation}

Next, we evaluate $\Phi(\rho^*,\brvq)$ for $\brvq > {\brvq}^*$.
Substituting the value of $\rho^*$ from \eqref{rhs} gives
\begin{equation}
  \log(\rho^*\brvq + 1) = \log(\bmax)- \log(\bmax-\brvq)
\end{equation}
and
\begin{equation}
  \varn - \frac{\rho^*}{\rho^*\brvq + 1} = \varn - \frac{1}{\bmax} = -
  (1 - \sqrt{\K})^2 .
\label{vfr}
\end{equation}

Substituting \eqref{vfr} into the second term in \eqref{phr} and
applying Lemma~\ref{lm} gives
\begin{multline}
  \int^b_a \log(x + \varn - \frac{\rho^*}{\rho^* \brvq + 1}) f_{\bS_1
    \bS_1^{\dag}}(x) \ \diff x \\
  = \frac{1}{2} \K \log(\K) -
  (\K-1)\log(\sqrt{\K}-1) - \sqrt{\K} .
\end{multline}

Thus, for $\brvq > {\brvq}^*$,
\begin{multline}
  \Phi(\rho^*,\brvq) = \log(\bmax)- \log(\bmax-\brvq) \\ + \frac{1}{2} \K
  \log(\K) 
  - (\K-1)\log(\sqrt{\K}-1) - \sqrt{\K} \\- \log(w(\varn))
  +
  (\K-1) \log(1 - \frac{1}{\K} v(\varn)) + v(\varn) .
\end{multline}

Also, $\Phi(\rho^*,\brvq) = \B \log(2)$.  We can explicitly solve for
$\brvq$ as follows
\begin{multline}
  \brvq = \bmax(1 - 2^{-\B} [\exp\{\frac{1}{2}\K\log(\K) \\ - (\K-1)
  \log(\frac{\K\sqrt{\K} - \K}{\K - v(\varn)}) \\- \log(w(\varn)) +
  v(\varn) - \sqrt{\K} \}]) .
\label{bbm}
\end{multline}

To solve $\B^*$, which corresponds to ${\brvq}^*$ \eqref{bst}, we
substitute ${\brvq}^*$ in \eqref{bbm}.

\bibliographystyle{IEEEtran}
\bibliography{IEEEabrv,rvq}

\end{document}